\documentclass[runningheads]{llncs}
\usepackage{graphicx}
\usepackage{xspace}
\usepackage{comment}
\usepackage{booktabs}
\usepackage{multirow}
\usepackage{adjustbox}

\usepackage{amsmath}
\usepackage{amssymb}
\usepackage{aas_macros}

\usepackage{xcolor}
\usepackage[newfloat, frozencache]{minted} 
\definecolor{vargreen}{HTML}{00BB00}
\definecolor{dodgerblue}{HTML}{1e90ff}
\definecolor{funccolor}{HTML}{AA22FF}
\definecolor{numbercolor}{HTML}{666666}
\SetupFloatingEnvironment{listing}{name=Code,placement=htp}

\usepackage{url}

\newcommand{\lcdm}{\ensuremath{{\mathrm{\Lambda CDM}}}\xspace}

\newcommand{\corrfunc}{\texttt{Corrfunc}\xspace}
\newcommand{\openmp}{\texttt{OpenMP}\xspace}
\newcommand{\rmax}{\ensuremath{{{\mathcal{R}}_\mathrm{max}}}\xspace}
\newcommand{\rmin}{\ensuremath{{{\mathcal{R}}_\mathrm{min}}}\xspace}

\newcommand{\pimax}{\ensuremath{{\pi_\mathrm{max}}}\xspace}
\newcommand{\rmaxcubed}{\ensuremath{{\mathcal{R}^{3}_\mathrm{max}}}\xspace}
\newcommand{\rmaxsqr}{\ensuremath{{\mathcal{R}^{2}_\mathrm{max}}}\xspace}
\newcommand{\rminsqr}{\ensuremath{{\mathcal{R}^{2}_\mathrm{min}}}\xspace}

\newcommand{\xiofr}{\ensuremath{{\xi(r)}}\xspace}
\newcommand{\wprp}{\ensuremath{{w_p(r_p)}}\xspace}

\newcommand{\avxft}{\texttt{AVX512}\xspace} 
\newcommand{\avxftf}{\texttt{AVX512F}\xspace} 
\newcommand{\avx}{\texttt{AVX}\xspace}
\newcommand{\sse}{\texttt{SSE}\xspace}
\newcommand{\fallback}{\texttt{Fallback}\xspace}
\newcommand\norm[1]{\left\lVert#1\right\rVert}

\newcommand{\simd}{\texttt{SIMD}\xspace}
\newcommand{\simdlen}{\texttt{SIMDLEN}\xspace}

\newcommand{\scrptimes}{\ensuremath{\scriptstyle \times}}

\begin{document}
\title{\textsc{Corrfunc}: Blazing fast correlation functions with \avxftf \simd Intrinsics}

\titlerunning{Blazing fast correlation functions with \avxftf}
%
\author{Manodeep Sinha\inst{1}\orcidID{0000-0002-4845-1228} \and
Lehman Garrison\inst{2}\orcidID{0000-0002-9853-5673}}
\authorrunning{Sinha \& Garrison}
%
\institute{SA 101, Centre for Astrophysics \& Supercomputing, Swinburne
  University of Technology, 1 Alfred St., Hawthorn, VIC 3122, Australia,\\
  ARC Centre of Excellence for All Sky Astrophysics in 3 Dimensions
  (ASTRO 3D)\\
\email{msinha@swin.edu.au}\and
60 Garden St., MS-10, Harvard-Smithsonian Center for Astrophysics, Cambridge, MA 02138}
\maketitle              
%

\setcounter{footnote}{0}

\begin{abstract}
Correlation functions are widely used in extra-galactic astrophysics to
extract insights into how galaxies occupy dark matter halos and in cosmology to
place stringent constraints on cosmological parameters. A correlation function
fundamentally requires computing pair-wise separations between two sets of
points and then computing a histogram of the separations. \corrfunc is an
existing open-source, high-performance software package for efficiently
computing a multitude of correlation functions.
In this paper, we will discuss the \simd \avxftf kernels within \corrfunc,
capable of processing 16 floats or 8 doubles at a time. The latest manually implemented \corrfunc \avxftf kernels show a
speedup of up to $\sim 4\times$ relative to compiler-generated code for double-precision
calculations.  The \avxftf kernels show $\sim 1.6\times$ speedup relative to the
\avx kernels and compare favorably to a theoretical maximum of $2\times$. In addition, by pruning
pairs with too large of a minimum possible separation, we achieve a
$\sim 5-10\%$ speedup across all the \simd kernels. Such speedups highlight the
importance of programming explicitly with \simd vector intrinsics for complex
calculations that can not be efficiently vectorized by compilers. \corrfunc is
publicly available at \url{https://github.com/manodeep/Corrfunc/}.
\end{abstract}
\section{Introduction}
Dark matter halos are spatially distributed in the Universe based
on the values of the cosmological parameters in the \lcdm cosmological model.
Galaxies live in dark matter halos, but how these galaxies populate halos depends on a
complex interplay between various astrophysical processes. We constrain this
`galaxy-halo connection' by statistically
comparing the spatial distribution of observed and modeled galaxies.
One such commonly used statistical measure is the correlation function.

A correlation function is the measure of the excess probability of finding a
pair of galaxies at a certain separation, compared to that of an Poisson
process. The simplest correlation function is a 3-D spatial one --- \xiofr:
\begin{equation}
dP = n_g(r) \left [1 + \xi(r)\right] dV,
\end{equation}
where $dP$ is the excess probability of finding a pair of galaxies, $n_g(r)$ is
the number density of galaxies, and $\xiofr$ is the correlation function.
In practice, to calculate the correlation function, we need to count the number of
galaxy-pairs found at a different separations. The separations themselves are
typically discretized as a histogram; as such, calculating a
correlation function amounts to calculating pairwise separations followed by a
spatial distance histogram.  We then
need to compare these pair counts with the the number expected from randomly distributed points for the same histogram bins. \xiofr is frequently calculated with the following\cite{peebles1980}:
\begin{equation}\label{eqn:xi}
    \xi(r) = \frac{N_{DD}(r)}{N_{RR}(r)} - 1,
\end{equation}
where $N_{DD}(r)$ and $N_{RR}(r)$ are respectively the number of ``data-data''
and ``random-random'' pairs in the histogram bin of separation $r+\delta r$.

In additional to the full 3-D separation $r$, the spatial separations can
be split into two components --- typically a line-of-sight ($\pi$) and a projected
separation ($r_p$). The line-of-sight direction is arbitrary but is usually chosen to be a coordinate axis.  When the separations are split into two components, the correlation function is computed as a 2D histogram of pair-counts, referred to as $\xi(r_p, \pi)$. The two-point projected correlation function, \wprp, is simply the integral of $\xi(r_p, \pi)$ along the line-of-sight and defined as:
\begin{equation}
\wprp  = 2 \int_0^{\pimax} \xi(r_p, \pi) d\pi
\end{equation}
For the remainder of the paper we will focus on these two correlation
functions --- \wprp and \xiofr.

A correlation function can be used to statistically compare any theoretically
generated set of mock galaxies to the observed galaxy clustering. Such a
comparison is frequently done within a Monte Carlo Markov Chain
process\cite{accurate_clustering}. For any reasonable MCMC estimates of the
posterior, we would need a large number of evaluations of the correlation
function. Hence a faster correlation function code is a key component for
cutting edge astrophysical research.

In \cite{corrfunc_paper} we showed that \corrfunc is {\em at least} $2\times$
faster than all existing bespoke correlation function codes. \corrfunc achieves
such high-performance by refining the entire domain into cells, and then
handling cell pairs with bespoke \simd kernels targeting various instruction
set architectures. In \cite{corrfunc_paper}, we presented three different
kernels targeting three different instruction sets -- \avx, \sse and the
\fallback kernels. In this work, we will present \avxftf kernels and additional optimizations.  

\subsection{Correlation Functions}
The simplest correlation function code can be written as:
\begin{listing}
\caption{\small{Naive code for a correlation function}}
\begin{minted}{C}
for(int i=0;i<N1;i++){
  for(int j=0;j<N2;j++){
    double dist = distance_metric(i, j);
    if(dist < mindist || dist >= maxdist){
      continue;
    }

    int ibin = dist_to_bin_index(dist);
    numpairs[ibin]++;
    weight[ibin] += weight_func(i, j);
  }
}
\end{minted}
\label{code:naivecorr}
\end{listing}

The only two components that are required to fully specify a correlation
function are:
\begin{itemize}
\item {\texttt{distance\_metric}}:  This specifies how to calculate the separation
  between the two points. For \xiofr, the separation is simply the {\it Euclidean} distance between the points
\item {\texttt{dist\_to\_bin\_index}}: This specifies how to map the separation
  into the histogram bin. Typically, there are $\sim 15-25$ bins logarithmically spaced
  between \rmin and \rmax, which span 2--3 orders of magnitude.
\end{itemize}

In this paper, we will be concentrating on two spatial correlation functions --- \xiofr
and \wprp. Consider a pair of distinct points, denoted by the subscripts $i$ and
$j$, with Cartesian positions $(x_i, y_i, z_i)$ and $(x_j, y_j, z_j)$. The
separations and the associated constraints (i.e., {\texttt{distance\_metric}})
for \xiofr and \wprp are:
\begin{align}
  \label{eqn:separations}
  \begin{split}
    \xiofr -
    \begin{cases}
      r := \sqrt{(x_i - x_j)^2 + (y_i - y_j)^2 + (z_i - z_j)^2} < \rmax, \\
    \end{cases}\\
    \wprp -
    \begin{cases}
      r_p := \sqrt{(x_i - x_j)^2 + (y_i - y_j)^2} < \rmax\\
      \pi := \lvert z_i - z_j \rvert  < \pimax. \\
    \end{cases}
  \end{split}
\end{align}
Thus, only pairs with $r < \rmax$ add to the histogram for \xiofr; while for
\wprp, pairs need to satisfy both conditions $r_p < \rmax$ and $\pi < \pimax$
before the histogram is updated. Note that the histogram for \wprp is still
only a 1-D histogram based off $r_p$; the $\pi$ constraint simply filters out
pairs with too large $\pi$ separation.

So far all we have is the histogram of pair-wise separations for the galaxies.
To fully evaluate the correlation function, we also need to evaluate the
histogram of pairs for randomly distributed points (see Eqn.~\ref{eqn:xi}). For
simple domain geometry, like a cubic volume with periodic boundary conditions as is common with cosmological simulation data, we can analytically
compute the number of random pairs for any histogram bin. Thus, we only need to compute the $N_{DD}(r)$ term in Eqn.~\ref{eqn:xi} to calculate \xiofr. A similar technique can be applied to \wprp as well. Consequently, we only need to compute the $N_{DD}(r)$ (i.e., an auto-correlation) term to calculate both \xiofr and \wprp.

In real galaxy surveys, the domain geometry is not so simple.  In angular extent, the domain is limited by foreground contamination and the sky area the telescope can see from its latitude; in radial extent, it is limited by the faintness of distant galaxies and other complex selection effects.  Thus, the $N_{RR}(r)$ and $N_{DR}(r)$ terms must often be computed by pair-counting; this is a major computational expense and is a motivating factor for our development of a fast pair-counting code.

\subsection{Partitioning the space into \rmax cells}\label{sec:grid}
For a given set of $N$ points, a naive implementation of a correlation function
would require evaluating all pair-wise separations and hence scale as
$\mathcal{O}(N^2)$. However, most correlation functions only require
calculating pair-separation up to some maximum separation \rmax. If we are only interested in pairs within \rmax, then computing the pairwise separation to {\em all} possible pairs only to discard majority of the computed separations is extremely computationally inefficient. We can create an initial list of potential neighbors and significantly reduce the total number of computations by first
partitioning the particles into cells of size at least \rmax. This idea of cell
lists\cite{Quentrec_cell_linked_list73}, or chaining meshes \cite{HOCKNEY1974148}, is used both in
molecular dynamics simulations\cite{gromacs_3_2001} and smoothed particle hydrodynamics
simulations\cite{swift_simd_verlet_2018}.

\begin{figure}
\centering
\includegraphics[width=0.55\linewidth,clip=true]{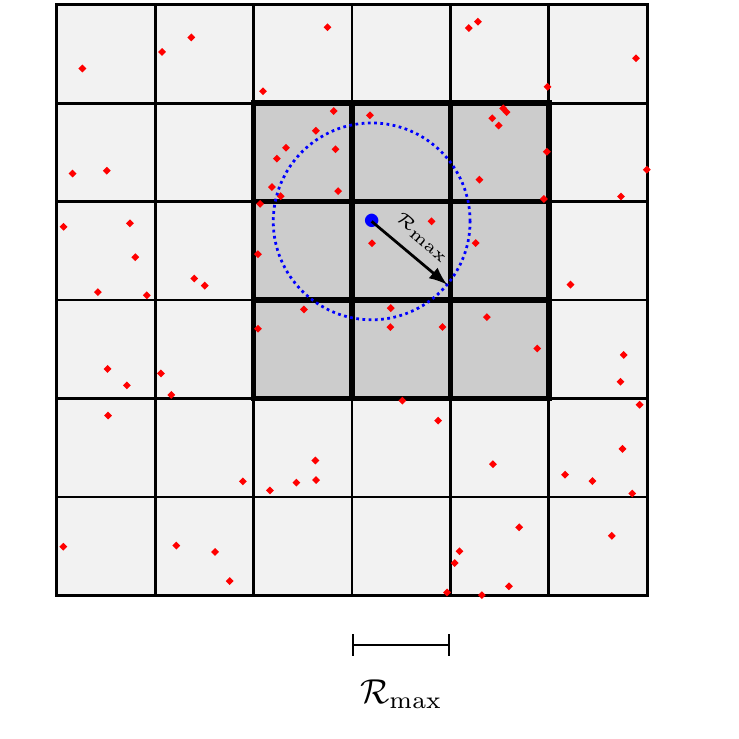}%
\caption{Partitioning the space to speed up the search for any potential pair
  within \rmax. The distribution of red points is gridded up on a lattice
  with cell-size at least \rmax. For any blue query point, all possible pairs
  involving the red points {\em must} lie within the 9 neighboring
  cells (dark gray shaded cells). With a similar lattice structure in 3 dimensions, we approximate a
  sphere of volume $\frac{4}{3}\pi\rmaxcubed$ with a cube of volume $27\rmaxcubed$. Figure adapted
  from \cite{corrfunc_paper}.}
\label{fig:grid}
\end{figure}

\begin{figure}
\centering
\includegraphics[width=0.55\linewidth,clip=true]{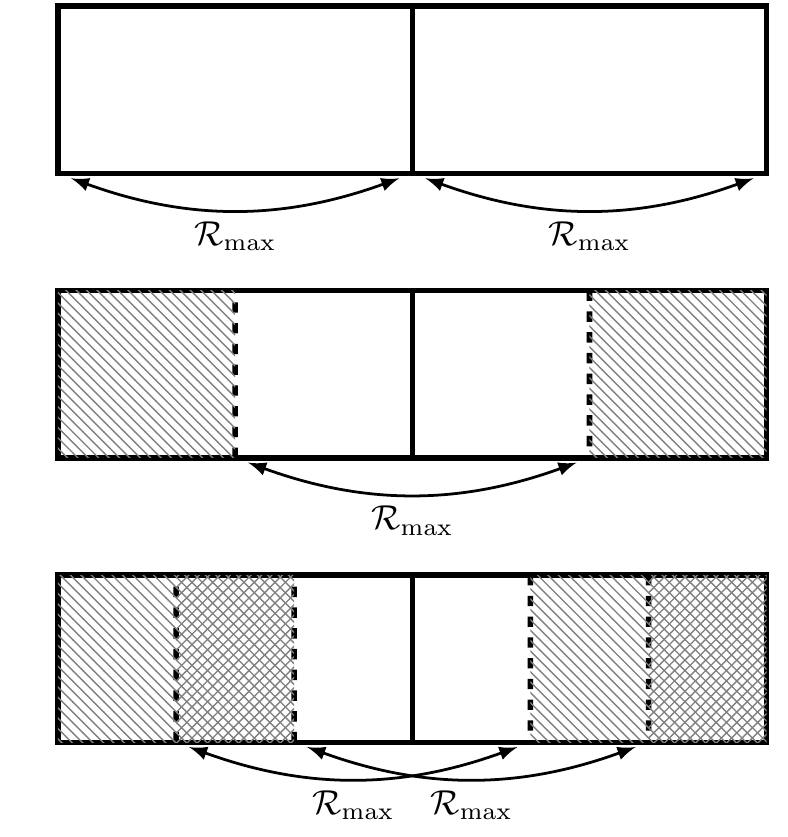}%
\caption{Refining the cell-size reduces the search volume. Particles in the
  shaded regions are separated by more than \rmax and can not have a
  pair. Compared to the top panel, the middle and lower panels need to inspect
  a smaller search volume. Figure adapted from \cite{corrfunc_paper}.}
\label{fig:grid_refine}
\end{figure}

After depositing the particles into cells of side at least \rmax, all possible pairs within
\rmax {\em must} lie within the neighboring cells. In Fig~\ref{fig:grid}, we
show the lattice of side \rmax imposed on the Poisson distributed red
points. For any given query point (in blue), we first locate the target cell
and then we immediately know all potential cells that may contain pairs.
However, this lattice implementation approximates the volume of a sphere of radius
\rmax by that of a cube with side $3\times\rmax$. Thus, if we compute all
possible pair-separations, then only 16\% of the separations will be within
\rmax; the remaining 84\% will be spurious\cite{cell_refinements_gonnet_07}. In
Fig.~\ref{fig:grid_refine}, we show that sub-dividing the cells further reduces
the effective search volume. Sub-dividing the cells into size $\rmax/2$, the spurious calculations drop to
$\sim 63\%$\cite{cell_refinements_gonnet_07}. Continuing to sub-divide further reduces the spurious
calculations even more, but the overhead of searching over many more neighboring cells
starts to dominate. In our experience with \corrfunc, we have found that bin sizes in the vicinity of $\rmax/2$ produce the fastest run-times for a broad range of use-cases.

\section{Overview of \corrfunc}
\subsection{Optimization Design}
\corrfunc provides a variety of high-performance, \openmp parallel, user-friendly correlation function codes. We have presented \corrfunc in \cite{corrfunc_paper} but to make this paper self-contained, we will briefly describe the design and optimizations implemented in \corrfunc.

\subsubsection{Creating 3D cells to reduce the search volume}
As we discussed in Sec~\ref{sec:grid}, the computational cost of searching for potential
neighbors can be reduced by dividing the space into 3D cells. Given two sets
of positions, the correlation function routines in \corrfunc
first computes a bounding box for the two datasets. The second step is to
sub-divide the space into 3D cells such that all possible pairs are located
within the neighboring cells (see Fig.~\ref{fig:grid}). However, for an optimal
grid size, we need to account for the separation in the specific correlation
function. As we showed in Eqn.~\ref{eqn:separations}, the \xiofr calculation
only needs pairs that satisfy $r < \rmax$ and correspondingly the grid-size is
some fraction of \rmax. For \wprp, we need two different separations -- $r_p$
and $\pi$, where $r_p$ is the separation between a pair of points in the
$X$-$Y$ plane, and $\pi$ is the separation along the $Z$ axis. Therefore, the
optimal grid-size is some fraction of \rmax in both $X$ and $Y$ axes, and
\pimax along the $Z$ axes. We combine the two cases into $\pi < \pimax$,
with the understanding that $\pimax = \rmax$ for \xiofr calculations.

\subsubsection{Improving cache utilization}
Within each cell, the particle positions are stored as a Structure-of-Array (SoA) variable.
The individual $X, Y, Z$
positions are copied from the input arrays and duplicated within dedicated
pointers in the SoA cell-array. Since we can load in the positions from the SoA
cell-array directly into vector registers without any shuffle operations, the
SoA operation is very conducive to vectorization.

\subsubsection{Reducing the search volume by sorting the $z$ positions}
With some appropriate sub-divisions, we can now locate all possible pairs that
satisfy the separation constraints among the neighboring cell pairs. For
\xiofr calculations, the spherical search volume of $4/3\pi\rmaxcubed$ is then
approximated with $27\rmaxcubed$. The particle positions stored in the SoA
are always sorted along the $Z$ axis. With such a sorting, we only need to seek $\pm \pimax$ from any particle position to find all possible pairs. Thus, with the $Z$-sorting, we reduce the search
volume along the $Z$ axis from $3\pimax$ to $2\pimax$.

\subsubsection{Reducing the number of cell pairs}
Once both the datasets have been gridded up and the particles assigned to
cells, we associate all possible pairs of cells that may contain a pair of
points within \rmax. The fundamental unit of work within \corrfunc involve such
pairs of cells.

For cases where the two datasets are distinct (cross-correlations), there are
no symmetry conditions to exploit. However, when the two datasets are identical
(auto-correlations), we can take advantage of symmetry. Only unique pairs of
cells need to calculated, and as long as we double the total number of pairs
found at the end, we will still have the correct number of pairs.
Both \xiofr and \wprp are auto-correlations and benefit from this optimization.

\subsection{Computing the minimum possible separation}\label{sec:min_sep_opt}
After culling for cell-pairs based on symmetry, we are left with cell-pairs
that are within \rmax along any one dimension, but might still be too far apart
when the full 3D separation is considered. To remove such cell-pairs, we need
to compute the minimum possible separation between these two cells. Based on
the linear offset between the cell-pairs along each dimension, we know the
minimum possible separations, $\Delta_X, \Delta_Y$ and $\Delta_Z$ along each
axis. However, these quantities only capture the extent of the individual cells
and do not reflect the actual positions of the particles within the
cell. Therefore, we also store the bounding box info for each cell. With the
bounding box, we can then compute the minimum possible separation
between the two cells by simply taking the
difference of the bounding axes along each dimension and then using
Eqn.~\ref{eqn:separations}. If the minimum possible separation is larger than \rmax, then
there {\em can not} be any valid pairs between the particles in the two cells
and we reject that cell-pair.

If a cell-pair passes this check, then there {\em
  may} exist at least one valid pair between the cells. So far the $\Delta_X$
and related quantities only reflect the minimum possible separation between the
cell-edges. $\Delta_X$ could really reflect the minimum separation between {\em
  any pair} of particles. Since we have the bounding box info for the secondary cell, we
increase each one of the three $\Delta_X$ quantities by the separation between
the secondary bounding box and the nearest secondary cell-edge. If most of the
secondary particles are concentrated towards the center of the secondary cell,
then we would increase $\Delta_X$ by a large amount and correspondingly prune a
larger chunk of the secondary particles.

\subsubsection{Late-entry and early-exit from the $j$-loop}\label{sec:loopcond}
For any pair of cells, we know the minimum possible particle separation
along each of the $X$, $Y$ and $Z$ axes -- $\Delta_x, \Delta_y$ and $\Delta_z$
respectively. We also store the positions for the X, Y and Z edges of the
primary cell nearest to the secondary cell -- $X_{edge}$, $Y_{edge}$ and
$Z_{edge}$ respectively. Since the minimum possible separation (between any
particle pairs) along $X$ and $Y$ axes is $\Delta_x$ and $\Delta_y$, and the
maximum total separation is \rmax, the maximum possible $dz:=(z_i - z_j)$ that
can satisfy $r < \rmax$:
\begin{align}
  dz_{\mathrm {max, all}} =  \sqrt{(\rmaxsqr - \Delta_X^2 - \Delta_Y^2)},
\end{align}
This $dz_{\mathrm {max, all}}$ only makes sense for \xiofr; for \wprp calculations
$dz_{\mathrm {max, all}}$ equals \pimax.

In addition, we can also compute the minimum possible separation between a
given primary particle and {\em any} secondary particle. We can make an
improved estimate for the minimum separation
between the $i'th$ primary particle and any secondary particle by using the $X$
and $Y$ positions of the primary particle. For every cell-pair we can then compute two
conditions Therefore, we can compute the minimum possible
\begin{align}
   dx_{i, \mathrm {min}} &= \Delta_X + \lvert (x_i - X_{edge}) \rvert,\\
   dy_{i, \mathrm {min}} &= \Delta_Y + \lvert (y_i - Y_{edge}) \rvert,
\end{align}
\begin{align}
dz_{i, \mathrm {max}} =  \sqrt{\rmaxsqr - dx^2_{i, \mathrm {min}} - dy^2_{i, \mathrm  {min}}}
\end{align}

Recall that the $z$ values in are sorted in increasing order, i.e., $z_j \le z_{j+1}
\le z_{j+2}, ...$, as well as $z_i \le z_{i+1} \le z_{i+2}, ...$. If we define
$dz_{ji} := z_j - z_i$, then the $dz$ values are also
in increasing order for a fixed value of $i$. Therefore, if any particle in the
second dataset has $dz_{ji} > dz_{i, \mathrm {max}}$, {\em all} future particles
from the second data also must have $dz > dz_{i, \mathrm {max}}$. When we encounter such a $j$
particle, we terminate the $j$-loop and continue on with the next iteration of
the outer $i$-loop.

Since the $dz_{ji}$ values are effectively sorted in increasing order, for a
given $i$, the smallest value of $dz_{ij}$ (note flipped subscripts) will occur for the final $j$-th
particle. Since the $z$ positions in the first dataset are also sorted in
increasing order, therefore $dz_{ij} \le dz_{(i+1)j} \le dz_{(i+2)j}...$
Thus, $dz_{i(N2-1)}$ is also the smallest possible value for
all remaining pairs between the two datasets. Therefore, if $dz_{i(N2-1)}$
exceeds $dz_{\mathrm {max, all}}$, no further pairs are possible and we can exit the
outer $i$-loop altogether.

For any $i$-th particle to have a pair in the second dataset, the condition
$ dz_{ji} < \lvert \operatorname{min}(\pimax, dz_{i, \mathrm {max}})\rvert$ must be met. Therefore
if $dz_{ji} >= dz_{i, \mathrm {max}}$, there can not be any possible pair between
this $i$-th particle and {\em any} $j$ particles. However, a different particle
from the first dataset might still be a valid pair with the remaining particles
in the second dataset. Therefore, we simply continue on with the next iteration
of the $i$-loop.

Since the $z$ positions are sorted, we continue to loop over the $j$ particles,
until we find a $j$ such that $dz_{ji} > - dz_{i, \mathrm {max}}$. This $j$ marks
the beginning particle for calculating pairwise separations.
\begin{listing}
m\caption{\small Late Entry and early exit condition}
\begin{minted}[escapeinside=||]{C}
  const |\textcolor{vargreen}{\textbf{DOUBLE}}| *zstart = z2;
  const |\textcolor{vargreen}{\textbf{DOUBLE}}| *zend = z2 + N2;
  const |\textcolor{vargreen}{\textbf{DOUBLE}}| dz|$_\mathrm{max,all}$| = |$\sqrt{\rmaxsqr - \Delta_X^2 - \Delta_Y^2}$|;
  for(int64_t i=0;i<N1;i++) {
    const |\textcolor{vargreen}{\textbf{DOUBLE}}| xpos = *x1++;
    const |\textcolor{vargreen}{\textbf{DOUBLE}}| ypos = *y1++;
    const |\textcolor{vargreen}{\textbf{DOUBLE}}| zpos = *z1++;

    |\textcolor{vargreen}{\textbf{DOUBLE}}| this_dz = *z2 - zpos;
    if(this_dz >= dz|$_\mathrm{max,all}$|) continue;

    const |\textcolor{vargreen}{\textbf{DOUBLE}}| dx = |$\Delta_X \;\ne 0 \; ? \; (\Delta_X + \norm{\mathrm{xpos - X_{edge}}})$|:0;
    const |\textcolor{vargreen}{\textbf{DOUBLE}}| dy = |$\Delta_Y \;\ne 0 \; ? \; (\Delta_Y + \norm{\mathrm{ypos - Y_{edge}}})$|:0;
    const |\textcolor{vargreen}{\textbf{DOUBLE}}| dz = |$\Delta_Z \;\ne 0 \; ? \; (\Delta_Z + \norm{\mathrm{zpos - Z_{edge}}})$|:0;
    const |\textcolor{vargreen}{\textbf{DOUBLE}}| sqr_sep = dx|$^2$| + dy|$^2$| + dz|$^2$|;
    if(sqr_sep >= |$\rmaxsqr$|) continue;
    const |\textcolor{vargreen}{\textbf{DOUBLE}}| dz|$_\mathrm{i,max}$| = |$\sqrt{\rmaxsqr-\mathrm{dx}^2-\mathrm{dy}^2}$|;

    while( (z2 < zend) && (*z2 - zpos) <= -dz|$_\mathrm{max,all}$|) {
      z2++;
    }
    if(z2 == zend) break;

    int64_t j = z2 - zstart;
\end{minted}
\label{code:loopcond}
\end{listing}

\subsubsection{Vector intrinsics in existing \simd kernels}\label{sec:simd_kernels}
In \cite{corrfunc_paper}, we presented the dedicated \avx and \sse
kernels. These kernels operate on cell pairs, finding all possible pairs
between the two cells and updating the histogram appropriately. Both the \avx
and \sse kernels have an associated header file each that maps C-macros to the correct
underlying intrinsic for both double and single precision floats. With such an
approach, the same lines of (pseudo-)intrinsics in the \simd kernels can be
seamlessly used for both single and double precision floats.

We vectorize the $j$-loop over the second set of points and process particles in
chunks of \simdlen; where \simdlen is 8 and 4 for single-precision \avx and
\sse respectively. For double precision calculations, \simdlen is halved and
equals 4 and 2 for \avx and \sse respectively.

Immediately following from Code~\ref{code:loopcond}, we can start processing
pairs of particles with \simd intrinsics.
\begin{listing}
\caption{\small The loop over secondary particles in \simd kernels}
\begin{minted}[escapeinside=||]{C}
  int64_t j = z2 - zstart;
  |\textcolor{vargreen}{\textbf{DOUBLE}}| *localz2 = z2;
  |\textcolor{vargreen}{\textbf{DOUBLE}}| *localx2 = x2 + j;
  |\textcolor{vargreen}{\textbf{DOUBLE}}| *localy2 = y2 + j;
  const |\textcolor{vargreen}{\textbf{SIMD\_DOUBLE}}| simd_xpos = |\textcolor{funccolor}{\textbf{SIMD\_SPLAT}}|(xpos);
  const |\textcolor{vargreen}{\textbf{SIMD\_DOUBLE}}| simd_ypos = |\textcolor{funccolor}{\textbf{SIMD\_SPLAT}}|(ypos);
  const |\textcolor{vargreen}{\textbf{SIMD\_DOUBLE}}| simd_zpos = |\textcolor{funccolor}{\textbf{SIMD\_SPLAT}}|(zpos);

  const |\textcolor{vargreen}{\textbf{SIMD\_DOUBLE}}| simd_sqr_rmax = |\textcolor{funccolor}{\textbf{SIMD\_SPLAT}}|(|$\rmaxsqr$|);
  const |\textcolor{vargreen}{\textbf{SIMD\_DOUBLE}}| simd_sqr_rmin = |\textcolor{funccolor}{\textbf{SIMD\_SPLAT}}|(|$\rminsqr$|);

  for(;j<=(N2 - |\textcolor{numbercolor}{\textbf{SIMDLEN}}|);j+=|\textcolor{numbercolor}{\textbf{SIMDLEN}}|) {
    const |\textcolor{vargreen}{\textbf{SIMD\_DOUBLE}}| simd_x2 = |\textcolor{funccolor}{\textbf{SIMD\_LOAD}}|(localx2);
    const |\textcolor{vargreen}{\textbf{SIMD\_DOUBLE}}| simd_y2 = |\textcolor{funccolor}{\textbf{SIMD\_LOAD}}|(localy2);
    const |\textcolor{vargreen}{\textbf{SIMD\_DOUBLE}}| simd_z2 = |\textcolor{funccolor}{\textbf{SIMD\_LOAD}}|(localz2);

    localx2 += |\textcolor{numbercolor}{\textbf{SIMDLEN}}|;
    localy2 += |\textcolor{numbercolor}{\textbf{SIMDLEN}}|;
    localz2 += |\textcolor{numbercolor}{\textbf{SIMDLEN}}|;

    const |\textcolor{vargreen}{\textbf{SIMD\_DOUBLE}}| dx = |\textcolor{funccolor}{\textbf{SIMD\_SUB}}|(simd_x2 - simd_xpos);
    const |\textcolor{vargreen}{\textbf{SIMD\_DOUBLE}}| dy = |\textcolor{funccolor}{\textbf{SIMD\_SUB}}|(simd_y2 - simd_ypos);
    const |\textcolor{vargreen}{\textbf{SIMD\_DOUBLE}}| dz = |\textcolor{funccolor}{\textbf{SIMD\_SUB}}|(simd_z2 - simd_zpos);

    if( |\textcolor{funccolor}{\textbf{ALL}}|(dz <= -dz|$_\mathrm{i,max}$|) ) continue;
    if( |\textcolor{funccolor}{\textbf{ANY}}|(dz >= dz|$_\mathrm{i,max}$|) ) j = N2;

    const |\textcolor{vargreen}{\textbf{SIMD\_DOUBLE}}| rp_sqr = |\textcolor{funccolor}{\textbf{SIMD\_ADD}}|(|\textcolor{funccolor}{\textbf{SIMD\_MUL}}|(dx, dx), |\textcolor{funccolor}{\textbf{SIMD\_MUL}}|(dy, dy));
    const |\textcolor{vargreen}{\textbf{SIMD\_DOUBLE}}| r_sqr = |\textcolor{funccolor}{\textbf{SIMD\_ADD}}|(rp_sqr, |\textcolor{funccolor}{\textbf{SIMD\_MUL}}|(dz, dz));

    const |\textcolor{vargreen}{\textbf{SIMD\_MASK}}| rmax_pairs = |\textcolor{funccolor}{\textbf{SIMD\_CMP\_LT}}|(r_sqr, sqr_rmax);
    const |\textcolor{vargreen}{\textbf{SIMD\_MASK}}| rmin_pairs = |\textcolor{funccolor}{\textbf{SIMD\_CMP\_GE}}|(r_sqr, sqr_rmin);
    |\textcolor{vargreen}{\textbf{SIMD\_MASK}}| pairs_left = |\textcolor{funccolor}{\textbf{SIMD\_AND}}|(rmax_pairs,rmin_pairs);
    if( |\textcolor{funccolor}{\textbf{NONE}}|(pairs_left) ) continue;

    /* histogram update here */
  }

  for(;j<N2; j++) { /* remainder loop */
  ...
\end{minted}
\label{code:simd}
\end{listing}
Since we can only process multiples of \simdlen with the \simd intrinsics, we
need an additional remainder loop to process any remaining particles in the
second dataset.

Once we have a set of \simd vectors, calculating the separations is
trivial. Note that we avoid the expensive {\texttt sqrt} operation in
Eqn.~\ref{eqn:separations} and always perform comparisons with squared
separations. Once we have the (squared) separations, we
we can create vector masks for separations that satisfy the
conditions in Eqn.~\ref{eqn:separations}. If no particles satisfy the distance
constraints, then we continue to process the next \simdlen chunk.
If there are particles that do satisfy all distance criteria, then the
histogram needs to be updated.

\subsubsection{Updating the histogram of pair-counts}
Within the \simd kernels, we have the array containing the values of the lower
and upper edges of the histogram bins. To update the histogram, we need to
first ascertain which bin any given $r$ falls into. The simplest way to locate the bin is to
loop through the bins, and stop when $r$ is within the bin-edges. Recall that
we avoid computing $r:=\sqrt{r^2}$, so the comparison is $r^2_{low} \le r^2 <
r^2_{hi}$. However, we can take advantage of how the typical bins are spaced
and perform the histogram update faster.
\begin{listing}
\caption{\small Histogram Update in \simd kernels}
\begin{minted}[escapeinside=||]{C}
for(int kbin=nbin-1;kbin>=1;kbin--) {
    const |\textcolor{vargreen}{\textbf{SIMD\_DOUBLE}}| m1 = |\textcolor{funccolor}{\textbf{SIMD\_CMP\_GE}}|(r2, rupp_sqr[kbin-1]);
    const |\textcolor{vargreen}{\textbf{SIMD\_MASK}}| bin_mask = |\textcolor{funccolor}{\textbf{SIMD\_AND}}|(m1, pairs_left);
    npairs[kbin] += |\textcolor{funccolor}{\textbf{POPCNT}}|(|\textcolor{funccolor}{\textbf{SIMD\_TEST}}|(bin_mask));
    pairs_left = |\textcolor{funccolor}{\textbf{SIMD\_CMP\_LT}}|(r2, rupp_sqr[kbin-1]);
    if( |\textcolor{funccolor}{\textbf{NONE}}|(pairs_left) ) break;
}
\end{minted}
\label{code:histupdate}
\end{listing}

Typically the histogram bins  are logarithmically spaced, and consequently, the outer
bins encompass a much larger volume than the inner bins. Therefore, many more
pairs of points are likely to fall in the outer bins than the inner
ones. Following \cite{Chhugani2012}, we loop backwards through the
histogram bins (see Code~\ref{code:histupdate}).
Within the \simd kernels, we create a mask that evaluates to
`true' separations that fell within a bin. We then run a hardware \texttt{popcnt} operation to count the number of bits set and update that particular
histogram bin. This iteration over the histogram stops once we have accounted
for all valid separations.

\subsection{Overview of \avxftf}
\avxft is the latest generation of instruction set architecture supported on
both the Intel SkyLake-SP, Skylake-X and the Xeon Phi x2000 (Knights Landing)
processors. \avxft expands the vector length to 512-bytes (compared to the
256-bytes in \avx) and introduces an additional mask variable type. Every
instruction now comes with a masked variety where some elements can be masked
out and alternate values specified for those masked-out lanes. The dedicated
mask variable type can be directly cast into an \texttt{uint16\_t}.\footnote{For
  double precision calculations, the upper 8 bits of the mask are identically set to 0}

Since \avxft is composed of several distinct instruction sets, both current and upcoming, we have
only targeted the specific subset -- \texttt{AVX512-Foundation} (\avxftf). \avxftf is meant to be supported by all current and future processors with \avxft support\footnote{\texttt{
    AVX512CD} is meant to allow vectorization of histogram updates but our
    attempts at automatic vectorization have proved futile so far}.

\subsection{\avxftf kernel implementation}
Within the \avxftf kernel, we employ the same late-entry and early-exit
conditions discussed in Sec~\ref{sec:loopcond}. In this subsection, we will
describe the \avxftf kernel once the first possible particle in the $j$-loop is identified.

The masked operations are quite handy when coding up the \avxftf kernel. For
instance, typically the array lengths are not an exact multiple of the \simd
vector length. Therefore, at the end of each vectorized loop, there is always a
`remainder' loop (see Code~\ref{code:simd}) to process the elements that
were left over. With the masked operations in \avxftf, we can pad out the
`remainder' points to be an exact \simd vector length; and set the mask for
these padded points as `false' (see Code~\ref{code:avxftf_kernel}). All of the
subsequent processing, including the
`load' from memory, then uses this mask as one of the operands.  Since there is
no longer any `remainder-loop', the \corrfunc \avxftf kernels are automatically more compact.
With such a masked load, we can completely avoid the `remainder' loop. We
simply continue to update the mask variable -- \texttt{mask\_left} -- while
processing the \simdlen separations per iteration of the $j$-loop. In other
words, we used masked comparisons, where the input mask already contains the
\texttt{mask\_left}.

Most of the remaining sections of the \avxftf kernel follows
similarly to the previous kernels. The two significant updates in the \avxftf
kernels are that we have used the \texttt{FMA} operations where possible,
and instead of the hardware \texttt{popcnt} instruction,
we have used a pre-computed array containing the number of bits set.
\begin{listing}
\caption{\small The $j$-loop in \avxftf kernels}
\begin{minted}[escapeinside=||]{C}
const uint16_t pairs_left_float[] =  {0xFFFF,
                                      0x0001, 0x0003, 0x0007, 0x000F,
                                      0x001F, 0x003F, 0x007F, 0x00FF,
                                      0x01FF, 0x03FF, 0x07FF, 0x0FFF,
                                      0x1FFF, 0x3FFF, 0x7FFF};

const uint8_t pairs_left_double[] = {0xFF,
                                     0x01,0x03,0x07,0x0F,
                                     0x1F,0x3F,0x7F};

const int64_t n_off = z2 - zstart;
const int64_t Nleft = N2 - n_off;
|\textcolor{vargreen}{\textbf{DOUBLE}}| *localz2 = z2;
|\textcolor{vargreen}{\textbf{DOUBLE}}| *localx2 = x2 + n_off;
|\textcolor{vargreen}{\textbf{DOUBLE}}| *localy2 = y2 + n_off;
for(int64_t j=0;j<Nleft;j+=|\textcolor{numbercolor}{\textbf{AVX512\_SIMDLEN}}|) {
    |\textcolor{vargreen}{\textbf{AVX512\_MASK}}| pairs_left = (Nleft - j) >= |\textcolor{numbercolor}{\textbf{AVX512\_SIMDLEN}}| ?
                            pairs_left_DOUBLE[0]:pairs_left_DOUBLE[Nleft-j];
    ...
}
\end{minted}
\label{code:avxftf_kernel}
\end{listing}

\section{Results}\label{sec:results}
In this section we will show the performance results from the newest \avxftf
kernel and compare to the previous \avx, \sse and \fallback kernels within
\corrfunc.
To run the benchmarks in Sec~\ref{sec:bench_grid_refine} and Sec~\ref{sec:bench_min_sep_opt}, we have used the Intel C compiler suite 2018 (icc/2018.1.163)
on a dedicated node with the Intel Skylake 6140 cpus at the OzSTAR supercomputing Centre
(\url{https://supercomputing.swin.edu.au/ozstar/}). \corrfunc was compiled with
the compilation flag \texttt{-O3  -xhost -axCORE-AVX512  -qopenmp}. We used the
git commit hash \texttt{7b698823af216f39331ffdf46288de57e554ad06} to run these
benchmarks.

To run the benchmarks in Sec~\ref{sec:bench_rmax}, we used the Intel C compiler
(2017) with the same compiler flags.  The hardware setup was a dual socket
machine with two Intel Xeon Gold 6132 @ 2.60GHz, for 28 threads total.

\subsection{Comparing the performance with sub-divided cells}\label{sec:bench_grid_refine}
\begin{figure}
\centering
\includegraphics[width=0.7\linewidth,clip=true]{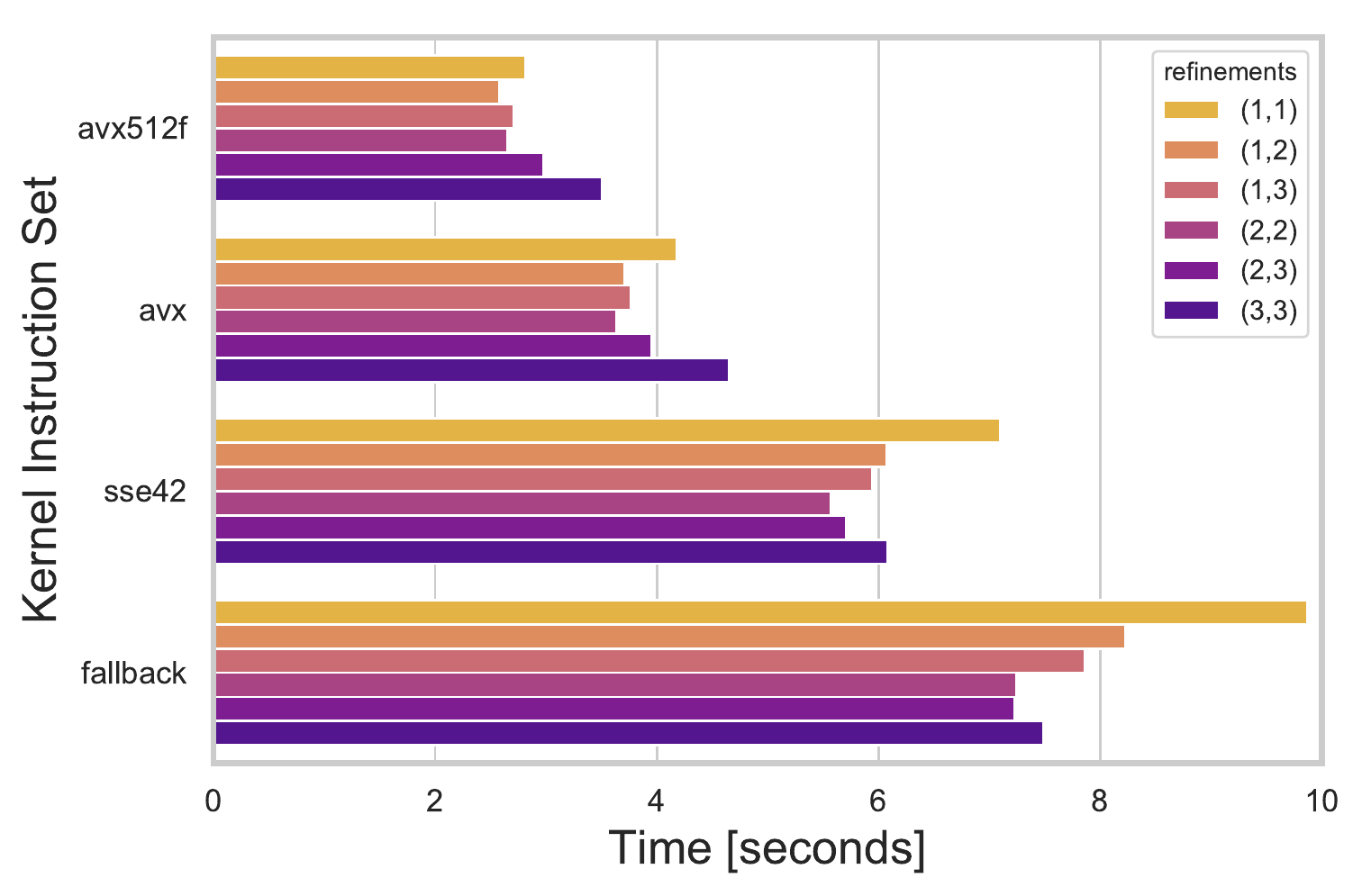}%
\caption{How sub-dividing the cells into sizes smaller than \rmax improves
  runtime performance. In this figure, the colors represent different
  combinations of the $X$ and $Y$ cell sizes ($dx$ and $dy$). The
  individual $(m, n)$ in the figure represent $(dx, dy) \gtrsim (\rmax/m, \rmax/n)$
  For example, $(1,1)$ is the case where $(dx, dy) \gtrsim \rmax$, $(1,2)$ is where $dx \gtrsim \rmax, dy  \gtrsim \rmax/2$ and so on. The \avxftf
  kernels are the fastest, followed by the \avx, the \sse and the \fallback
  kernels. There is variation of up to 50\% in runtime (for the \fallback case) for different \texttt{bin\_refine\_factors}; the variation in runtime is less pronounced ($\sim 20\%$) in the \simd kernels.  Thus, choosing an optimal cell-size is an important aspect of performance. }%
\label{fig:bin_refine_kernel}
\end{figure}

We showed in Section~\ref{sec:grid} that if we keep the cell sizes at
\rmax, then only 16\% of the pairs computed are likely to be necessary. Within
\corrfunc, we have the runtime option of refining the cell-size
further. There are 3 \texttt{bin\_refine\_factors} corresponding to each of the
$X$, $Y$ and $Z$ axes. These \texttt{bin\_refine\_factors} dictate how many
further sub-divisions of the initial \rmax cell are made.
Based on past experience, refining along the $Z$ axis only degrades
performance; therefore, we have fixed the $Z$ refinement at 1 and only allowed the
refinements to vary along the $X$ and $Y$ axes. In
Fig~\ref{fig:bin_refine_kernel}, we show how reducing the cell-size
impacts the total runtime for a \xiofr calculation, with \rmax=25 in a periodic
box of side 420. We
let the $X$ and $Y$ bin refinement factors to vary between 1 and 3. Every
combination of $(m, n)$ represents cell-sizes of $(dx, dy) \gtrsim (\rmax/m, \rmax/n)$.
In Fig.~\ref{fig:bin_refine_kernel} we see that \avxftf kernel is by far the
fastest, followed by the \avx, the \sse and the \fallback kernels. This
relative ordering of the kernels is in keeping with the relative ordering of
the vector register sizes.

Within each kernel group, the coarsest sub-division --- $(1, 1)$ --- is the
slowest with the performance improving with higher bin refinement
factors. However, all kernels are slower for bin refinements of $(3, 3)$
indicating that the overhead of looping over neighbor cells dominates over the
reduction in the search volume from the bin refinements (see
Fig~\ref{fig:grid_refine}). The improvement in the runtime for the \fallback kernel is drastic --- $\sim 50\%$ faster in moving from $(1,1)$ to $(2,3)$. The \simd kernels also show variations in the runtime with different \texttt{bin\_refine\_factors} but the scatter seems to reduce with wider vector registers. For instance, the \avxftf kernels show the least variation in runtime with the \texttt{bin\_refine\_factors} while the \sse kernels show the largest variation.

Within \corrfunc, we simply set the default \texttt{bin\_refine\_factor} to $(2, 2, 1)$
and recommend that the user experiment with various refinement factors to find
out what produces the fastest code for their use-case. With the new \avxftf
kernel, the same default bin refinement factors continue to be the fastest option.
As we showed in Fig~\ref{fig:grid_refine}, since the search volume reduces with
higher bin refinement factors, we do expect a reduction in runtime. However,
the exact speedup obtained is likely to be strongly dependent on \rmax and the
number density of the particles.  Exploring this dependence of the speedup on
\rmax and the particle load and automatically setting the
\texttt{bin\_refine\_factors} to close-to-optimal values would be a great improvement
to the \corrfunc package.

\subsection{Comparing the performance with bounding box optimizations}\label{sec:bench_min_sep_opt}
\begin{figure}
\centering
\includegraphics[width=0.7\linewidth,clip=true]{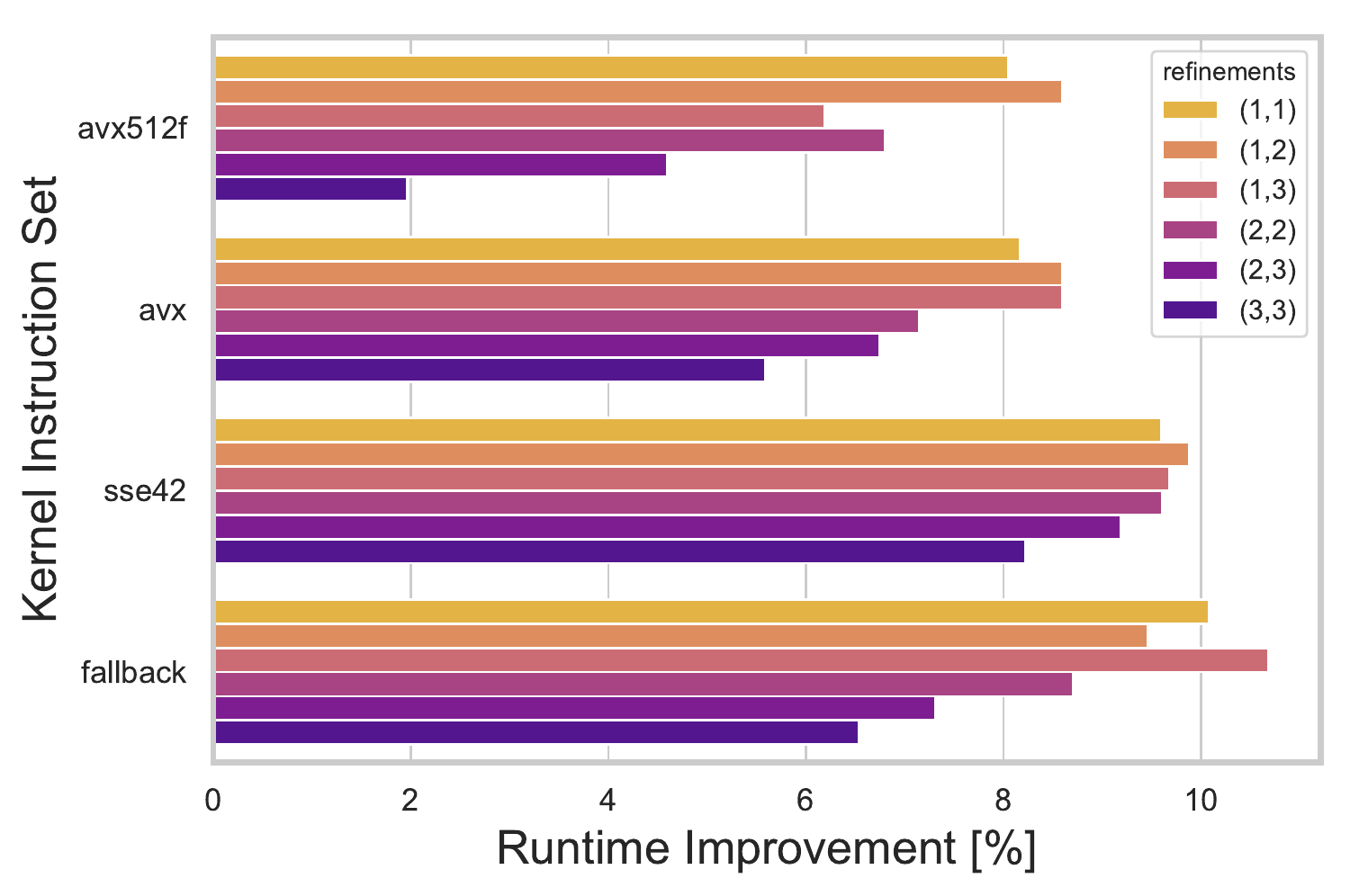}%
\caption{How the runtime changes for each \simd kernel by calculating the minimum separation
  between cell-pairs, as well as the minimum separation for a primary particle
  and all remaining particles in the secondary cell (see Section~\ref{sec:min_sep_opt}). The colors show
the different refinements (see Section~\ref{sec:grid} and Section~\ref{sec:bench_grid_refine}). The improvement ranges from
few\% to $\sim$ 10\% across all the \simd kernels; however, the performance
improvement seems to reduce when there are a larger number of sub-divisions.}%
\label{fig:min_sep_opt_simd_kernel}
\end{figure}

In Section~\ref{sec:loopcond}, we discussed how we maintain a bounding box
for every cell. After computing the minimum possible separation based on the
bounding box, we can then reject any cell-pair that can not contain a particle
pair within \rmax. In addition, for every primary particle, we can compute the
minimum possible separation to any secondary particle. If this minimum
separation is larger than \rmax, then we simply continue with the next primary
particle. In Fig.~\ref{fig:min_sep_opt_simd_kernel} we show the performance
improvement based on the minimum separation calculations. We can see that that
the performance improvement is typically in the 5-10\% range, with the latest
instruction sets showing the smallest improvement. Since the minimum separation
calculation is done in scalar mode, that optimization means more time is spent
in the scalar section of the \simd kernel and consequently the kernels with the
widest vector registers show the smallest improvement. In addition, we also see
that the improvement generally reduces as the number of sub-divisions
increases. This is most likely a combination of two effects -- i) increased runtime overhead
for processing larger number of neighbor cells, and ii) overall lower amount
of computation per cell pair (due to smaller number of particles per cell)
means the potential work avoided with the minimum separation calculation is
lower.

The dataset we have used for the benchmark contains a modest $1.2$ million
galaxies. It is quite likely that the minimum separation optimization will have
a larger impact for larger dataset (or equivalently, a larger \rmax).

\subsection{Comparing the performance of the \simd kernels}\label{sec:bench_rmax}
In the preceding section we saw that the \avxftf kernels in \xiofr are the fastest for a
for a fixed value of \rmax. We also wish to validate that \avxftf performs well for a broad range of \rmax.
Typical values of \rmax range from few percent to 25\% of
the periodic box size, so we explore this range in our tests. In the following,
the box size of the test dataset is 420; we show performance for \rmax values between 10 and 100.

In this section, we will compare the performance of the
four \simd kernels for a range of \rmax values, with the calculations being
done in double precision. For \xiofr, increasing \rmax
means increasing the search volume as $\rmaxcubed$. For \wprp, we have fixed
$\pimax = \rmax$; hence, increasing \rmax also increases the search volume by
$\rmaxsqr\times\pimax = \rmaxcubed$.  Thus, in both cases we expect an asymptotic $\rmaxcubed$
runtime scaling.

\begin{figure}
\centering
\includegraphics[width=\linewidth,clip=true]{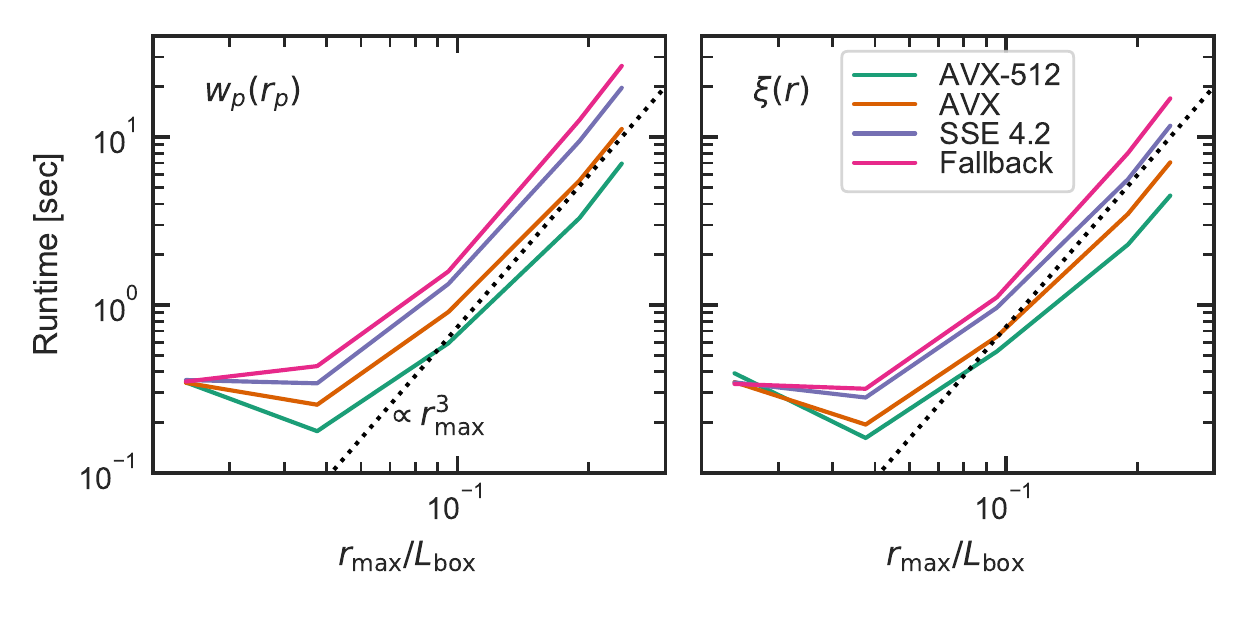}%
\caption{Correlation function runtime versus \rmax. For the \wprp
  calculations (left panel), we have set $\pimax:=\rmax$. The effective search volume is
  then either that of a sphere, $4/3 \pi \rmaxcubed$ for \xiofr or a cylinder $\pi
  \rmaxsqr \pimax = \pi \rmaxcubed$, since $\pimax = \rmax$ for this test. Thus, in both the \xiofr and \wprp cases, the
  search volume scales as $\rmaxcubed$ and correspondingly both the correlation
  functions scale accordingly. As we summarized in
  Table~\ref{table:simd_speedup_rmax}, we find that the \avxftf kernels are
  faster by $\sim 3.9\times$ relative to the \fallback kernel, and
  $\sim 1.6\times$ relative to the \avx kernel.}
\label{fig:rmax_scaling}
\end{figure}

In Fig.~\ref{fig:rmax_scaling}, we show
how the various kernel run-times scale.  We see the expected \rmaxcubed behavior at large \rmax, with the \avxftf kernels being the
fastest for reasonably large \rmax. At the lowest \rmax values, each cell
contains only a small number of particles and it is likely that there is not
sufficient computational work to keep the wider vector registers
full.\footnote{A low \rmax is potentially a case where the
 \texttt{bin\_refine\_factors} need to be set to $(1, 1, 1)$ to boost the particle
  occupancy in the cells}

Now that we have seen the raw performance, let us compare the speedups obtained
by the vectorized kernels relative to the \fallback kernel which contains no explicit SIMD instructions. Since the
calculations are done in double precision, a priori we expect a theoretical
maximum of $8\times, 4\times, 2\times$ speedup for the \avxftf, \avx and \sse
kernels, assuming the compiler is not able to automatically generate any vectorized instructions.
Even in the case that the compiler is able to do so, this test will measure how much more efficient
our vector implementation is than the compiler's.

In Table~\ref{table:simd_speedup_rmax} we show the speedup obtained with the
various \simd kernels relative to the \fallback kernel for a range of values
for \rmax. We see that for $\rmax = 10$, there is essentially no performance
boost from the vectorized kernels. Once \rmax is $\gtrsim 80$, the speedup seems
to stabilize at $\sim 3.8\times, 2.3\times, 1.3\times$ for the \avxftf, \avx
and the \sse kernels respectively. These achieved speedups are within a factor
of 2 of the theoretical maximum speedup. More interestingly, the \avxftf is
$1.6\times$ faster than the \avx kernels, compared to the theoretical maximum
of $2\times$ speedup from the wider vector registers. We also use the
\texttt{FMA} operations in \avxftf kernels, which also adds to the potential
speedup.
\begin{table}
\centering
\label{table:simd_speedup_rmax}
\caption{Speedup from the \simd kernels relative to the \fallback kernel as a
  function of \rmax. For \wprp calculations, we have set $\pimax = \rmax$. All
  of these calculations are done with a simulation box of periodic size $420.0$.}
\begin{adjustbox}{max width=\linewidth}
\begin{tabular}{ccccccccc}
\toprule
\multirow{2}{*}{$\boldsymbol{\rmax}$}        &
\multicolumn{4}{c}{$\boldsymbol{\wprp}$}     &
\multicolumn{4}{c}{$\boldsymbol{\xiofr}$}    \\
\cmidrule(l{1.0em}r{1.0em}){2-5}
\cmidrule(l{1.0em}r{1.0em}){6-9}
&
\multicolumn{1}{c}{$\boldsymbol{\avxftf}$}     &
\multicolumn{1}{c}{$\boldsymbol{\avx}$}        &
\multicolumn{1}{c}{$\boldsymbol{\sse}$}        &
\multicolumn{1}{c}{$\boldsymbol{\fallback}$}   &
\multicolumn{1}{c}{$\boldsymbol{\avxftf}$}     &
\multicolumn{1}{c}{$\boldsymbol{\avx}$}        &
\multicolumn{1}{c}{$\boldsymbol{\sse}$}        &
\multicolumn{1}{c}{$\boldsymbol{\fallback}$}   \\
\midrule
  10.0 &          1.1\scrptimes &          1.0\scrptimes &          1.0\scrptimes &          1.0\scrptimes &          1.0\scrptimes &          1.0\scrptimes &          0.9\scrptimes &          1.0\scrptimes \\
  20.0 &          2.7\scrptimes &          1.8\scrptimes &          1.3\scrptimes &          1.0\scrptimes &          2.2\scrptimes &          1.8\scrptimes &          1.1\scrptimes &          1.0\scrptimes \\
  40.0 &          3.0\scrptimes &          1.8\scrptimes &          1.2\scrptimes &          1.0\scrptimes &          2.4\scrptimes &          1.9\scrptimes &          1.2\scrptimes &          1.0\scrptimes \\
  80.0 &          3.9\scrptimes &          2.3\scrptimes &          1.3\scrptimes &          1.0\scrptimes &          3.6\scrptimes &          2.3\scrptimes &          1.4\scrptimes &          1.0\scrptimes \\
 100.0 &          3.8\scrptimes &          2.4\scrptimes &          1.4\scrptimes &          1.0\scrptimes &          3.8\scrptimes &          2.4\scrptimes &          1.5\scrptimes &          1.0\scrptimes \\

\bottomrule
\end{tabular}
\end{adjustbox}
\end{table}

\section{Conclusions}
In this paper, we have presented \avxftf kernels for calculating correlation
functions within the open-source package \corrfunc. These \avxftf kernels have
been manually coded with vector intrinsics and make extensive use of masked operations
to compute the separations and then update the histogram of pair-counts. The
\avxftf kernels show a typical speedup for $\sim 3.8\times$ relative to the
compiler-generated code within the \fallback kernel.  The speedup is $\sim
1.6\times$ relative to the \avx kernel and compares very well to the
theoretical maximum of $2\times$.  In addition, by efficiently pruning
pairs that have separations larger than \rmax, we gained up to a 10\% speedup. This paper and \cite{corrfunc_paper}
highlight the importance of combining domain knowledge, efficient algorithms, and dedicated vector
intrinsics for complex calculations. Such combinations are particularly powerful when the underlying problem is difficult for the compiler to efficiently vectorize, as is the case for \corrfunc.

\section{Acknowledgements}
MS was primarily supported by NSF Career Award (AST-1151650) during main \corrfunc
design and development. MS was also supported by the Australian Research
Council Laureate Fellowship (FL110100072) awarded to Stuart Wyithe and by funds for the
Theoretical Astrophysical Observatory (TAO). TAO is part of the All-Sky Virtual
Observatory and is funded and supported by Astronomy Australia Limited,
Swinburne University of Technology, and the Australian Government.  The latter
is provided though the Commonwealth's Education Investment Fund and National Collaborative
Research Infrastructure Strategy (NCRIS), particularly the National eResearch
Collaboration Tools and Resources (NeCTAR) project. Parts of this research were
conducted by the Australian Research Council Centre of Excellence for All Sky
Astrophysics in 3 Dimensions (ASTRO 3D), through project number CE170100013.

%
%
%
\bibliographystyle{splncs04}
\bibliography{master}
\end{document}